\documentclass[aps,12pt,superscriptaddress,amsfonts,amssymb,amsmath]{revtex4}

\usepackage{graphicx}
\usepackage{epsfig}
\usepackage{makeidx}
\usepackage{epstopdf}

\begin{document}

\title{Common Origin of Power-law Tails \\ in Income Distributions and Relativistic Gases}

\author{G.\ Modanese \footnote{Email address: giovanni.modanese@unibz.it}}
\affiliation{Free University of Bozen-Bolzano, Faculty of Science and Technology, Bolzano, Italy}

\linespread{0.9}

\begin{abstract}

\bigskip

Power-law tails are ubiquitous in income distributions and in the energy distributions of diluted relativistic gases. We analyze the conceptual link between these two cases. In economic interactions fat tails arise because the richest individuals enact some protection mechanisms (``saving propensity'') which allow them to put at stake, in their interactions, only a small part of their wealth. In high-energy particle collisions something similar happens, in the sense that when particles with very large energy collide with slow particles, then as a sole consequence of relativistic kinematics (mass dilation), they tend to exchange only a small part of their energy; processes like the frontal collision of two identical particles, where the exchanged energy is 100\%, are very improbable, at least in a diluted gas.
We thus show how in two completely different systems, one of socio-economic nature and one of physical nature, a certain feature of the binary microscopic interactions leads to the same consequence in the macroscopic distribution for the income or respectively for the energy. 

\end{abstract}

\maketitle

\section{Introduction}

In the last years a growing attention has been devoted to statistical phenomena in physics, social sciences and natural sciences, in which the distribution function of the variable of interest (like energy, for a gas, or individual income, for a society) has a tail that is not exponential, with rapid decrease, but can be fitted by a power law. Such tails are often called ``fat tails'' and their existence has several consequences, since it means in practice that the events of the tail are not so rare, in comparison to the bulk of the distribution, as those in a Gaussian distribution. 

Examples of fat tails can be given in the most various subjects \cite{fat}. In income distributions, the presence of fat tails (called ``Pareto tails'' in that context) means that the super-rich in the society are more numerous than expected from log-normal statistics. In cosmic ray physics, a fat tail in the particle energy distribution signals the relative abundance of particles with very large energy, actually an energy many orders of magnitude larger than the average energy. 

It is known that in certain cases the kinetic equations of statistical mechanics, which can also be applied to economic phenomena like money exchange, correctly predict an exponential behaviour of the equilibrium distribution function \cite{ya1,ya2,sch}. 
It is then natural to ask in what conditions a power-law tail may arise, or in other words, what are the fundamental mechanisms generating fat tails. Many arguments have been proposed in the literature \cite{new}. Since fat tails are so ubiquitous, and make their appearance in systems whose dynamical laws seem to have nothing in common, it is unlikely that they all originate from just one dynamical process; but one could try to identify some leading conditions. 

\section{Saving propensity}

In Boltzmann statistical mechanics, an essential requirement for an exponential distribution appears to be the micro-reversibility of the interactions. But also kinematics plays a role: it has been shown that relativistic kinematics is compatible with distributions having a fat tail, called the Kaniadakis functions \cite{ka1}, which are obtained from the minimization of a suitable entropy functional. In econophysics models, Pareto tails  have been reproduced by introducing the concept of ``saving propensity''  \cite{ch0,cha,BM1}, provided the saving propensity is distributed in an heterogeneous way among individuals; there is also in this case a connection to reversibility, since an exchange between individuals having different saving propensity cannot be reversed. The Kaniadakis functions are a remarkably good fit for econophysics models, too  \cite{BM1}

\begin{figure}
\begin{center}
\includegraphics[width=10cm,height=7cm]{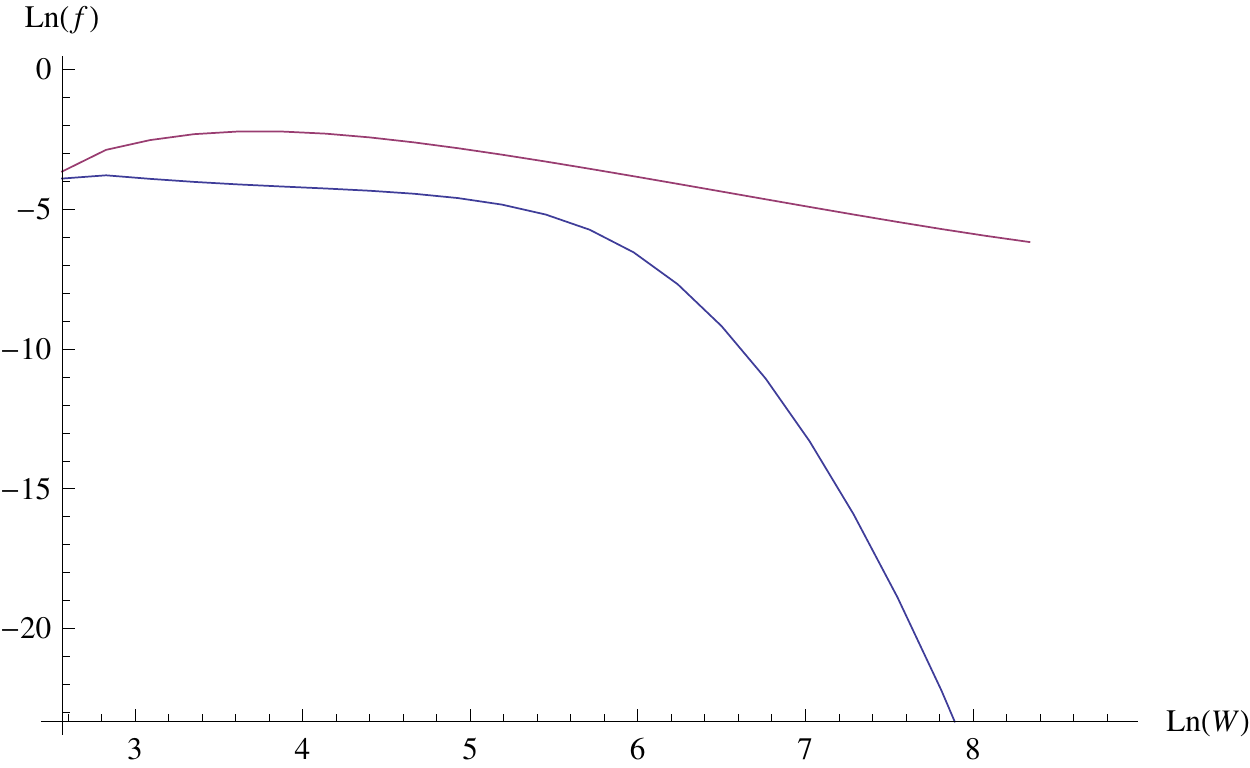}
\caption{Two examples of income distributions with power-law fat tail (upper curve) and without fat tail, i.e. with exponential decay (lower curve). These are obtained from the discretized kinetic theory of Ref.\ \cite{BM2}, with different  interaction parameters. $W$ denotes wealth or income, $f(W)$ the income distribution.} 
\label{f01}
\end{center}  
\end{figure}

The essential role of saving propensity in the formation of Pareto tails is clear also when the basic wealth exchange models are endowed with more complex structures. In \cite{BM2} we have considered a discretized kinetic theory of wealth exchange with taxation and redistribution effects depending on several parameters.
In this approach one supposes to re-group the $N$ agents into $n$ income classes, with $n \ll N$ and write a system of ordinary differential equations of the Boltzmann type which describe transitions between the classes. The fundamental variables are, in this case, the populations $x_i$ of the classes. The interactions are represented by terms quadratic and cubic in these variables. The asymptotic limit 
for $t \to \infty$ of the populations $x_i(t)$ gives the equilibrium income distribution. Fig.\ \ref{f01} shows an example of two income distributions with $n=25$ and two different sets of parameters. The upper distribution in the log-log plot exhibits a straight-line tail (corresponding to a power-law tail), while the lower distribution decreases much quicker at large income (exponential or ``leptokurtic'' tail). Note that the total income in the two cases is different. The upper curve is obtained with redistribution parameters $\tau_{min}=0.1$, $\tau_{max}=0.45$, $\theta=0.7$, while in the second curve redistribution is absent.

We would like now to show that there is a deep and interesting conceptual link between fat tails in income distributions and in the energy distributions of diluted relativistic gases, independently from specific dynamical features and issues of reversibility. The idea can be summarized as follows. In economic interactions fat tails arise because the richest individuals enact some ``protection mechanisms'' which allow them to put at stake, in their interactions, only a small part of their wealth, and protect the rest. This is, at least, a necessary condition. In high-energy particle collisions something similar happens, in the sense that when particles with very large energy collide with less energetic particles, they tend to exchange only a small part of their energy, so to make virtually impossible (at least in a diluted gas) processes like the classical frontal collision of two billiard balls, where the exchanged energy is 100\%. This happens as a sole consequence of relativistic kinematics, and quite independently from the dynamics. In other words, very fast particles tend to preserve most of their energy in collisions, much in the same way as very rich individuals tend to preserve most of their wealth in economic interactions.

\section{Effective mass and energy transfer in relativistic kinematics}

In this section we give a simplified estimate (for the present interdisciplinary purposes) of energy transfer in relativistic collisions. Before that, let us recall some facts on the $\kappa$-distribution and the cosmic rays energy distribution.
In \cite{ka2,ka3} G. Kaniadakis has shown that the deformation of the distribution function introduced by the parameter $\kappa$ emerges naturally within Einstein's special relativity, so that one can see the $\kappa$-deformation as a pure relativistic effect. To this end, one first proves that the $\kappa$-deformed sum of the momenta of two particles is the additivity law for the relativistic momenta (the $\kappa$-sum is used in a generalized entropy minimization procedure leading to the $\kappa$-distribution). By considering, in the framework of special relativity, the $\kappa$-statistics of an ensemble of identical particles, one therefore arrives at a distribution with a power-law tail without any assumption on dynamics, but using only kinematics. This is what is needed to explain why the cosmic ray spectrum, which extends over 13 magnitude orders in energy, approx. from $10^8$ to $10^{20}$ eV, is not exponential but follows a power law $E^{-a}$, with exponent $a$ between 2.7 and 3.1. The further assumption is made, that the cosmic rays can be viewed as an equivalent statistical system of identical particles with masses near the proton mass $m_p$. The resulting fits of the $\kappa$-deformed distribution display an excellent agreement with real data. It is finally possible to determine the value of $\kappa$ in relation to $m_p$, $c$, the absolute temperature $T$ and the Boltzmann constant $k_B$. For a recent review on the relativistic mechanism generating the power law tails in statistical distributions in high energy physics see also \cite{ka4}.

\begin{figure}
\begin{center}
\includegraphics[width=10cm,height=7cm]{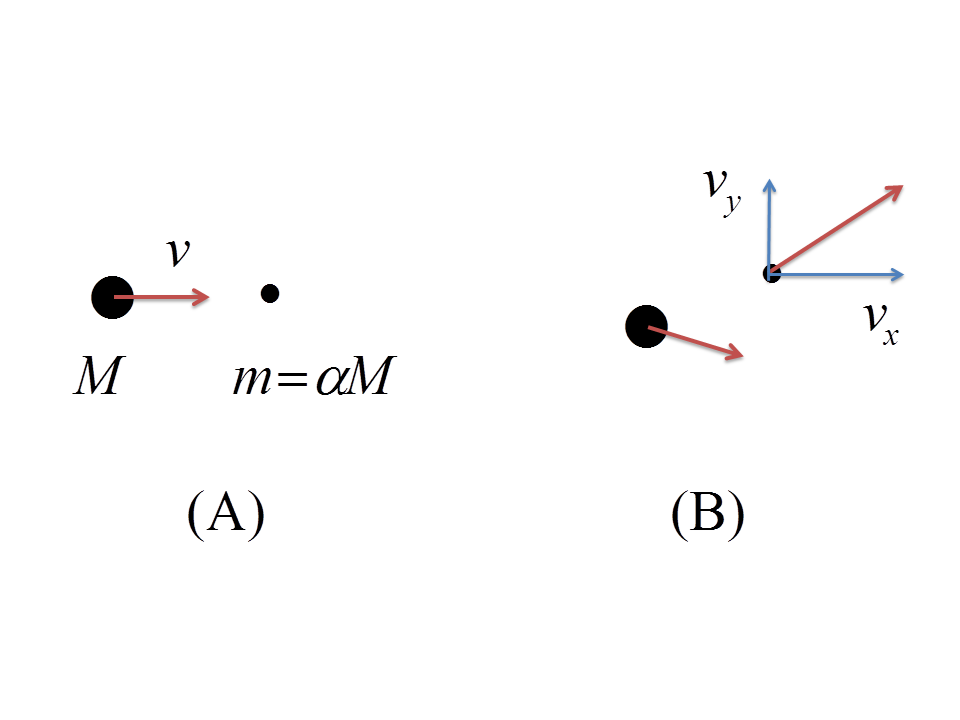}
\caption{Notation for the scattering between an incoming particle of mass $M$ and a target particle at rest of mass $\alpha M$, where $\alpha \ll 1$. (A): Before collision. (B): After collision. The ratio $R_{max}$ between the maximum energy transferred in the collision and the energy of the incoming particle (corresponding to the economic saving propensity) turns out to be $R_{max} \simeq 2\alpha$ (eq.s (\ref{ee1}) - (\ref{ee5})).} 
\label{f02}
\end{center}  
\end{figure}

Let us now consider at first a non-relativistic scattering between two particles, such that the mass of the target particle is $\alpha$ times the mass of the incoming particle ($\alpha \leq 1$). Suppose that the incoming particle has initial velocity $v$ in the $x$ direction. The computation starts from the conservation laws of energy and momentum, which we can write in the form
\begin{equation}
\frac{1}{2}M{v^2} = \frac{1}{2}M\left( {V{'_x}^2 + V{'_y}^2} \right) + \frac{1}{2}m\left( {{v_x}^2 + {v_y}^2} \right)
\label{ee0}
\end{equation}
\begin{equation}
Mv = MV{'_x} + m{v_x}; \qquad MV{'_y} + m{v_y} = 0
\label{ee01}
\end{equation}
Through some straightforward steps, the quantities $V{'_x}$ and $V{'_y}$ can be completely eliminated from these equations. Then consider the transversal component $v_y$, after the scattering, of the velocity of the target (Fig.\ \ref{f02}). The condition of energy conservation allows for a solution only if
\begin{equation}
v_y \leq \frac{v}{\alpha+1}
\label{ee1}
\end{equation}
Supposing that $\alpha$ is small (incoming particle much heavier than the target), it follows that approximately
\begin{equation}
v_y \leq v \qquad \qquad {\rm for} \ \  \alpha \ll 1
\end{equation}
The ratio $R$ between the energy transferred to the target and the initial energy of the incoming particle is
\begin{equation}
R=\frac{\alpha}{v^2} \left(v_x^2+v_y^2 \right) 
\end{equation}
where $v_x$ is the final velocity of the target; therefore inserting the maximum value of $v_y$, we have
\begin{equation}
R_{max} \simeq \alpha \left(\frac{v_x^2}{v^2}+1 \right)  \qquad \qquad {\rm for} \ \  \alpha \ll 1
\end{equation}
But with the maximum value of $v_y$, energy conservation gives $v_x=v/(\alpha+1)$, which is very close to $v$ for small $\alpha$, and in conclusion
\begin{equation}
R_{max} \simeq 2\alpha \qquad \qquad {\rm for} \ \  \alpha \ll 1
\label{ee5}
\end{equation} 
This ratio, which corresponds to the saving propensity, is therefore approximately equal to twice the mass ratio.

Let us then proceed to the relativistic case through an heuristic reasoning. We first observe that the rule of momentum conservation in a scattering between two particles is equivalent to Newton's law of action and reaction. According to the latter, when there is an interaction between a massive particle and a light particle, the acceleration of the light particle is much larger (inversely proportional to mass). This is for instance the reason why, in agreement with eq.s (\ref{ee1}) - (\ref{ee5}), when a truck hits a bycicle, it cannot pass to the bycicle an appreciable fraction of its energy. Now, if we consider a relativistic scattering between two identical particles, with the target particle at rest in the laboratory system, to a first approximation we can assume that there is a Newtonian force acting between the particles, and that the effective mass of the particle in motion is larger than the other, due to the relativistic mass dilation factor $\gamma=(1-v^2/c^2)^{-1/2}$. Therefore, also in this case the force will ``displace'' the target particle more than the incoming particle, thus preventing in most cases a large transfer of energy.

It is possible to check in a simple way that for a one-dimensional collision the relativistic mass dilation effect holds not only for small velocities, when $\gamma$ is close to 1, but also when $\gamma$ is very large. The relativistic second law of dynamics in the presence of a constant force reads
\begin{equation}
{\bf F} = \frac{{d{\bf{P}}}}{{dt}} = \frac{d}{{dt}}\frac{{m{\bf{v}}}}{{\sqrt {1 - \frac{{{v^2}}}{{{c^2}}}} }} = \frac{d}{{dt}}(m\gamma {\bf{v}})
\label{ee6}
\end{equation}
Computing the derivative one obtains
\begin{equation}
{\bf F} = m\gamma {\bf{\dot v}} + m{\gamma ^3}{\bf{v}}\left( {\frac{{{\bf{v}} \cdot {\bf{\dot v}}}}{{{c^2}}}} \right)
\label{ee7}
\end{equation}

Now suppose that the motion occurs only in the $x$ direction and denote by $v$ the $x$ component of ${\bf v}$ and by $F$ the $x$ component of ${\bf F}$. Then eq.\ (\ref{ee7}) reduces to 
\begin{equation}
F=\left( {\gamma  + {\gamma ^3}\frac{{v^2}}{{{c^2}}}} \right)m{{\dot v}}
\label{ee8}
\end{equation}
In the limit when the velocity of a particle is close to $c$, we see that its effective mass is amplified not only by a factor $\gamma$, but by a factor of the order of $\gamma^3$.

The second law is only an approximation for the full scattering, as can be seen from the fact that in principle the full relativistic conservation laws of energy and momentum for identical particles do allow as a limit case a frontal collision with 100\% transfer (a possibility which does not exist in the classical scattering (\ref{ee1}) - (\ref{ee5}), when the incoming mass is larger than the target mass). The phase space volume available for these extreme processes, however, is small, as can be checked through more complex mathematical arguments and as is ultimately proven by the presence of the power-law tail in the Kaniadakis distribution, which is only based on statistics and relativistic kinematics, and so in practice on relativistic phase space. The elementary argument given above holds for rarefied gases, in which the only relevant interactions involve couples of particles. For fat tails in plasmas, which originate in general in a more complicated way, see for instance \cite{space}.

An alternative argument for electromagnetic interactions in a material can be patterned after the derivation of the Bethe-Bloch equation. Consider a fast particle of charge $q$ interacting with a stationary particle of charge $Q$, with impact parameter $b$. The maximum force (either attractive or repulsive) and an effective interaction time are approximated as follows:
\begin{equation}
{F_{\max }} = \frac{{Qq}}{{4\pi {\varepsilon _0}{b^2}}}, \qquad \Delta t = \frac{b}{v}
\label{ee-r1}
\end{equation}
The impulse delivered to the stationary particle is mostly transverse, resulting in a momentum transfer
\begin{equation}
\Delta p \simeq {F_{\max }}\Delta t = \frac{{Qq}}{{4\pi {\varepsilon _0}bv}}
\label{ee-r2}
\end{equation}
The corresponding energy transfer is
\begin{equation}
\Delta E = \frac{{\Delta {p^2}}}{{2m}} \propto \frac{1}{{{v^2}}}
\label{ee-r3}
\end{equation}
showing that fast particles lose less energy per unit distance in a material than do slow particles.

\section{Conclusions}

The presence of fat tails in cosmic rays distributions, in spite of the variety of dynamical processes involved, is a strong indication that a general kinematical mechanism is at work. In much the same way, the ubiquity of Pareto tails in the income distribution of different countries in different epochs shows that the mechanism behind it is independent from the detailed economic structure of society, and must come from a universal feature like the heterogeneous saving propensity. In this work we have established a connection between these two facts.

We have shown how in two completely different systems, one of socio-economic nature and one of physical nature, a certain feature of the binary microscopic interactions leads to the same consequence in the macroscopic distribution for the income or respectively for the energy.

For the two systems the logical argument proceeds in different ways. 

(a) In the economic models the saving propensity is introduced ``by hand'': one writes exchange equations which contain a heterogeneous saving propensity factor chosen ad hoc. These equations respect a conservation law for the total money, which can be seen as the equivalent of energy conservation in physical collisions; however, there is no equivalent of momentum conservation. The outcome of the model, i.e. the macroscopic distribution, is obtained through the numerical solution of systems of ordinary differential equations, in our approach, or through stochastic differential equations or agent-based simulations, in other approaches. 

(b) For the relativistic gas, one starts from first principles, in the form of exact conservation equations without any arbitrary assumption -- except for the simplification of considering only binary collisions. Then, analytical techniques of relativistic statistical mechanics lead to a macroscopic distribution with a power-law tail. In order to clarify this result we have made the point that just supposing that the particles interact through a certain fixed force, the relativistic variation of their effective mass implies a tendency of the more energetic particles to lose less energy in the collisions.
 
In spite of the limited scope of the simplified models used, it is known that Pareto tails appear everywhere in economics, and fat tails are ubiquitous in plasmas and in gases which are far from ideal. While it is impossible to prove economic facts based on physical facts, the analogy we have pointed out can suggest in which direction we should look in order to see new effects. This has happened frequently in applications from physics to economics (for instance, one often looks for ``phases'', ``transitions'', ``critical points'' in economic systems). In some cases, however, empirical facts established in economics could have analogues in physics. It has been observed, for instance, that the fraction of the total population found in the Pareto tail changes dramatically when the economy expands or contracts due to external interactions or variations in productivity \cite{Sil}. It would be interesting to build a proper model of this phenomenon in economics in the first place, and then to look for physical analogues.

\end{document}